\documentclass[aps,prl,twocolumn,groupedaddress]{revtex4-1}

\usepackage{graphicx}
\usepackage{txfonts}

\begin{document}


\title{Thermal light edge enhancement ghost imaging of phase objects}


\author{Hanquan Song, Zhidan Yuan, Yuhang Ren, Dayu Zhao, Zili Zhang, Zhiyuan Zheng, and Lu Gao$^*$}

\address{
School of Science,
\\ China University of Geosciences, Beijing 100083, China\\
$^*$Corresponding author: gaolu@cugb.edu.cn}


\date{\today}

\begin{abstract}
We propose a experimental scenario of edge enhancement ghost imaging of phase objects with nonlocal orbital angular momentum (OAM) phase filters. Spatially incoherent thermal light is separated into two daughter beams, the test and reference beams, in which the detected objects and phase filters are symmetrically placed, respectively. The results of simulation experiment prove that the edge enhanced ghost images of phase objects can be achieved through the second-order light field intensity correlation measurement owing to the OAM correlation characteristics. Further simulation results demonstrate that the edge enhanced ghost imaging system dose not violate a Bell-type inequality for the OAM subspace, which reveals the classical nature of the thermal light correlation.
\begin{description}
\item[Usage]
Manuscript.
\item[PACS numbers]
 (030.0030) Coherence and statistical optics; (110.0110) Imaging systems; (070.0070) Fourier optics and optical signal processing; (060.5060) Phase modulation.
\end{description}
\end{abstract}


\maketitle

\section{I. INTRODUCTION}
Ghost imaging is a nonlocal image acquisition technique in which an image can be reconstructed by
using a light beam that never interacts with the object through intensity correlation measurement. The phenomenon seems intriguing but can be physically explained by quantum mechanical or classical correlation with a two-photon entangled source or a thermal light source, respectively \cite{shih}-\cite{chengjing}. The spatial correlation properties of a quantum entangled source and a thermal light source are different. The quantum entangled source behaves as a mirror, whereas the classical thermal source looks like a phase-conjugate mirror in the correlated imaging \cite{cao}.
The debate on the quantum vs classical nature of ghost imaging has lead to other interesting two-photon imaging effects using classical sources \cite{resch}, e.g., computational ghost imaging \cite{shapiro}\cite{altmann}, sub-shot noise ghost imaging of weak absorbing objects \cite{brida}\cite{samantaray}, x-ray region ghost imaging \cite{pelliccia}\cite{hanshen} and even wavelength transforming ghost imaging of a single photon imaging system \cite{morris}\cite{aspden}.

Recently, the nature of the correlations between different OAM components dependent on the phase of the fluctuations of pseudothermal light source have been studied, and the research results show that the thermal light correlation effects share similarities with the quantum correlations in the azimuthal degree of freedom \cite{omar}\cite{lugao}. Furthermore, several investigations have addressed two-photon correlation in regard to the high dimensional OAM structure \cite{mair}-\cite{torres}. Ghost imaging and object identification explorations with OAM quantum correlations have been implemented \cite{patarroyo}-\cite{zheyang}. Among these studies, B. Jack et al. of the University of Glasgow \cite{jack} applies the edge enhancement techniques in classical imaging to ghost imaging and proves that a phase filter, nonlocal with respect to the object, leads to enhanced coincidence images. They also interpret the high-contrast ghost images as a violation of a Bell inequality to demonstrate its quantum nature of the implementation of ghost imaging.

Here in this letter we propose an experimental scenario to accomplish the edge enhancement imaging of phase objects by making use of nonlocal phase filter. The intrinsic of the experiment is owing to the OAM correlation of the incoherent thermal light source. Although there are many studies on the thermal light correlation, we have not yet found the relevant exact proof of its intrinsic to our knowledge. Furthermore, we apply the proposed edge enhance ghost imaging system to measure the Bell-type violation. A circle phase object with $\pi$-phase step along the circumference and four phase filter with different orientations are used to achieve the result of non-violation of the Clauser-Home-Shinomy-Holt Bell-type inequality. The study results provide an explicit proof of the nature of the thermal light ghost imaging system embedded with edge enhancement imaging for phase objects.
\section{II. RESULTS}
\begin{figure}[htbp]
\centering \includegraphics[width=9cm]{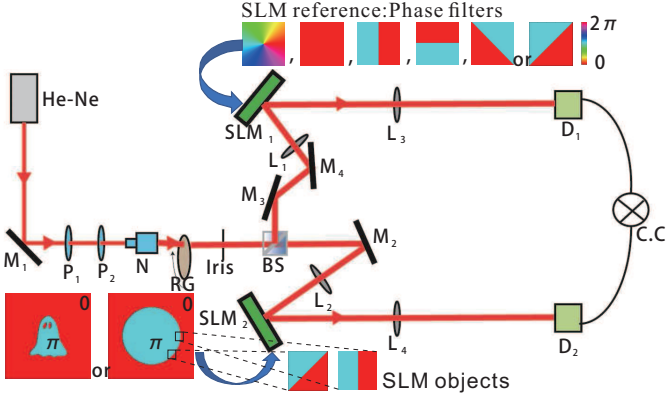}
\caption{ The schematic diagram of the thermal light edge enhancement ghost imaging of phase objects. The phase object is
stepped across the transverse plane of the $SLM_{2}$ and then imaged onto the $CCD$. At each position the local form of the object can be treated as a simple phase step of appropriate orientation and correlation measurements are made for different reference holograms. The details of the experimental schematic are explained in the main text.}
\label{Fig.1}
\end{figure}

The scenario scheme is shown in Fig.1.  A He-Ne laser is used to illuminate coherent light beam to a slowly rotation glass($RG$) to produce pseudo-thermal light beam. $N$ is a telescope used to collimate and expand the speckle size of the laser beam, while polarizers $P_{1}$ and $P_{2}$ can modulate the intensity of the laser beam. The spatially incoherent light beam are generated after the $RG$ and then it is divided into two twin beams, a test beam and a reference beam, by a 50:50 beam splitter ($BS$). The phase object and phase filter are implemented by two $SLMs$, which are placed in each beam symmetrically. Both $SLMs$ are in the imaging plane of the thermal light source after the $RG$, and then re-imagined to the transverse planes of the $CCDs$ at the end of each light arm. $D_{1}$ and $D_{2}$ are connected with a data acquisition card embedded in a personal computer to implement the intensity correlation calculation. The resolution of $D_{1}$ and $D_{2}$ is 8.3 $\mu m$. Owing to the phase indistinguishability of the correlation characteristics of the incoherent thermal light, images of the detected phase objects can not be observed from the intensity distribution in the transverse plane of each detector or even from the correlation measurement between the light field intensity of the two detectors. However, a special class of the optical vortices carries OAM, which is characterized by an azimuthal phase dependence of the form
$\exp(il\phi)$. Here $l$ is the OAM mode number and $\phi$ is the azimuthal angle. Recent studies \cite{omar}\cite{lugao} have demonstrated
the nature of the correlations between different OAM components dependent on the phase of the fluctuations of pseudothermal light source.
 Thus the edge enhancement effect of the pure phase object can be implemented through second-order light field intensity correlation measurement.
 The phase object and phase filter holograms are loaded on the $SLMs$ symmetrically in the test and reference beams, which are shown in the inner
 pictures of Fig. 1. The size of the phase object is $500\ast500$ pixels and the phase filter hologram is $10\ast10$ pixels. The red parts of the
 inner pictures represent the phase of $0$, and the blue part of the pictures represent the phase of $\pi$. The second-order correlation function
 between the two separated light beam can be given by $\Delta G^{(2)}(l_{t},l_{r})\propto |\int d\phi_{t} exp[i\phi_{t}(l_{t}-l_{r})]|^{2}$, where $l_{t}$ and $l_{r}$ characterize the helical phase modes of the test and reference beams, and $\phi_{t}$ presents the phase distribution of the test beam. The second-order correlation exactly happens with $l_{t}=l_{r}$, which will decrease with the increase of the difference between $l_{t}$ and $l_{r}$. Any object, or part thereof, can be described by an appropriate superposition of the modes carrying a phase term of $\exp(il\phi)$. If the phase object contains the same OAM component as the phase filter, the edge enhancement effect can be accomplished through correlation measurement.


The phase object "ghost" is loaded on $SLM_{2}$ in the test beam, which is shown in Fig.2 (a). And the phase filter with topology charge of $l_{r}=0$, $l_{r}=1$ or a binary grating with a $\pi$-phase step in different directions is loaded on $SLM_{1}$ in the reference beam. The image is acquired by stepping the object in the transverse plane and recording the second-order correlation measurement value. The computational simulation results are shown in Fig. 2.

When a spiral phase filter, with index $l_{r}$, is placed in the reference arm, the correlation measurement result is proportional
to the model component of the object corresponding to $l_{t} = l_{r}$. The intensity distribution of the test beam is shown in Fig.2 (b) and no information of the object is contained owing to the incoherent characteristics of the thermal light beam. When the phase filter of $l_{r}=0$ is placed on $SLM_{1}$, any part of the object with uniform phase of $0$ gives a high correlation value owing to $l_{t}=l_{r}$ and the correlation value comes to zero for other part with the non-zero phase distribution. The edge enhancement image is shown in Fig.2 (c), in which the correlation value at the edge of the phase object is lower than other parts with uniform phase distribution of $0$. Then the phase filter on $SLM_{1}$ is replaced by $l_{r}=1$. For a part of the object with a $\pi$-phase step, an expansion in terms of exp($il_{obj}\phi$) contains non-zero components for $l_{obj}=\pm1$ \cite{jack}. Such a phase step therefore gives a high second-order correlation measurement value for $l_{ref}=\pm1$. The simulation result is shown in Fig.2 (d), in which the edge of the phase object gives higher second-order correlation value than other parts of the object. Fig.2 (e) and (f) show the partial edge enhancement ghost images with oriented phase filters of horizontal or vertical $\pi$-phase step distributions, respectively. Correlation images appear where the contrast of the edge detection depends on the relative orientation of the edge with respect to the reference phase step.

\begin{figure}[htbp]
\centering \includegraphics[width=8.0cm]{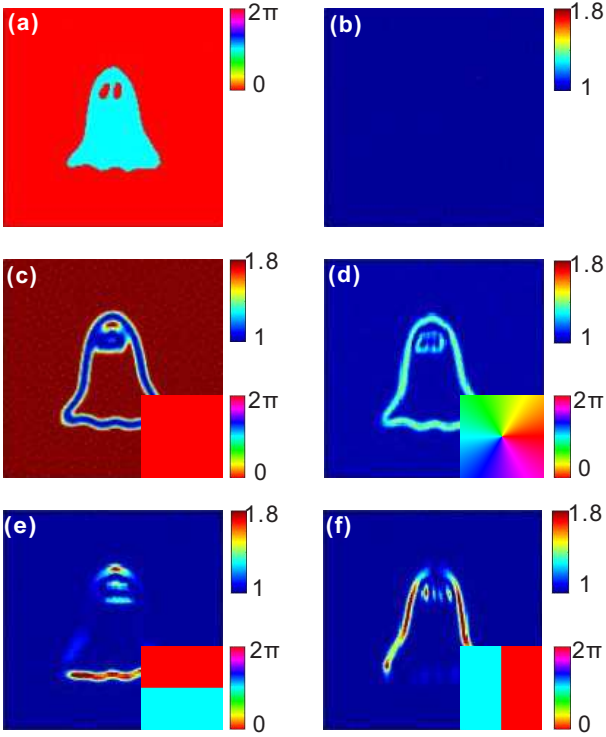}
\caption{Simulation results of the thermal light edge enhancement ghost imaging experiment. (a) Phase object. (b) The intensity distribution in the plane of $D_{2}$. (c)-(f) Edge enhanced ghost images of the phase object shown in (a). The phase filter holograms in the reference beam are shown in insets. }
\label{Fig.2}
\end{figure}

Bell inequality is a powerful method of discriminating whether a physical phenomenon is of quantum signature or not. Though the thermal light correlation effects share similarities with the quantum correlations both in the spatial and azimuthal degree of freedom, however, the correlated images obtained with thermal light have a finite background. The thermal light correlation characteristics will not violate the Bell inequality, but no explicit proof has been given previously \cite{jack}. Here we provide the categorical demonstration through computational simulation based on the thermal light edge enhanced ghost imaging scenario. A circular phase object is chosen to replace the "ghost" phase object in the test beam and the phase filters in the reference beam have the form of the $\pi$-phase step with different orientations. Both the phase object and the phase filter can be generated by holograms loaded on the $SLMs$, which are shown in Fig. 1. Owing to the $\pi$-phase step distribution of the circular phase object and phase filter, the measurement of the Bell-type inequality is based on the state space for the OAM states $l_{ref}=\pm1$ with an equally weighted superposition. Only if the phase filter has the same azimuthal orientation as the phase circular object, the enhanced ghost image appears to be "bright" through thermal light second-order correlation measurement. Since the circular phase object is much larger than the phase filter, the enhanced ghost image is achieved by stepping the phase object in its transverse plane and combining the results of each time correlation measurement. The orientations of the azimuthal phase filter in the reference beam are $0$, $\frac{\pi}{2}$, $\frac{\pi}{4}$, and $\frac{3\pi}{4}$, respectively. We record the thermal light second-order correlation value as a function of the relative angle between the phase step edge of the circular phase object and that of the phase filter in the reference beam. The simulation experimental results are shown in Fig. 3.

Such a Bell inequality measurement can be quantified by making use of  a Clauser-Home-Shimony-Holt Bell type inequality with the $|S|\leq2$, and
\begin{eqnarray}
S=E(\theta_{A},\theta_{B} )-E(\theta_{A},\theta_{B}^{\prime })+E(\theta_{A}^{\prime},\theta_{B} )+E(\theta_{A}^{\prime},\theta_{B}^{\prime}), \nonumber\\
\end{eqnarray}
where $\theta_{A}$ present the orientation of filter and $\theta_{B}$ present azimuth angle of the phase object\cite{Lebedev}. $E(\theta_{A},\theta_{B})$ is calculated from the correlation measurement in particular orientations,
\begin{eqnarray}
E(\theta_{A},\theta_{B} )=\frac{C(\theta_{A}, \theta_{B} )+C(\theta_{A}^{\ast },\theta_{B}^{\ast } )-C(\theta_{A}^{\ast },\theta_{B} )-C(\theta_{A},\theta_{B}^{\ast })}{C(\theta_{A},\theta_{B} )+C(\theta_{A}^{\ast },\theta_{B}^{\ast } )+C(\theta_{A}^{\ast },\theta_{B} )+C(\theta_{A},\theta_{B}^{\ast })}, \nonumber\\
\end{eqnarray}
where $ \theta_{(A,B)}^{\ast}=\theta_{(A,B)}+\frac{\pi}{2} $,the subscripts A and B denote the orientation of filter and azimuth angle of the phase object,and C is the second-order correlation value according to FIG.3(e). Fig. 3(a)-3(d) present images of the circular phase object with the reference hologram oriented at $0$, $\frac{\pi}{2}$, $\frac{\pi}{4}$, and $\frac{3\pi}{4}$, respectively. Fig. 3(e) shows the variation curves calculated from the enhanced images in each of Fig. 3(a)-3(d). The image data in Fig. 3(e) are averaged from the area unit of 8 radial pixels by 3 azimuthal degrees, and sinusoidal patterns are shown from the second-order correlation measurement. It should be stated that the maximum values of the curves (a) and (b) are larger than those of the (c) and (d). In the condition of (c) and (d) the azimuthal phase filter is inclined of 45 degree, which has the zigzag distribution owing to the square shape of each pixel point of the hologram loaded on the $SLM$. The zigzag edge phase-step distribution of the phase filter will destroy the value of the second-order correlation measurement.

By submitting $\theta_{A}=0$, $\theta_{B}=\frac{\pi}{8}$, $\theta_{A}^{\prime}=\frac{\pi}{4}$ and $\theta_{B}^{\prime}=\frac{3\pi}{8}$ into Eqs. (1) and (2), The maximal violation of the Bell inequality can be obtained \cite{Leach}. In this condition we achieve the maximum value of S is 0.57623 through computational simulation, which is less than the local-hidden-variable bound of 2. The simulation result proves that the edge enhanced ghost imaging system is of the classical signature.

\begin{figure}[htbp]
\centering \includegraphics[width=8.0cm]{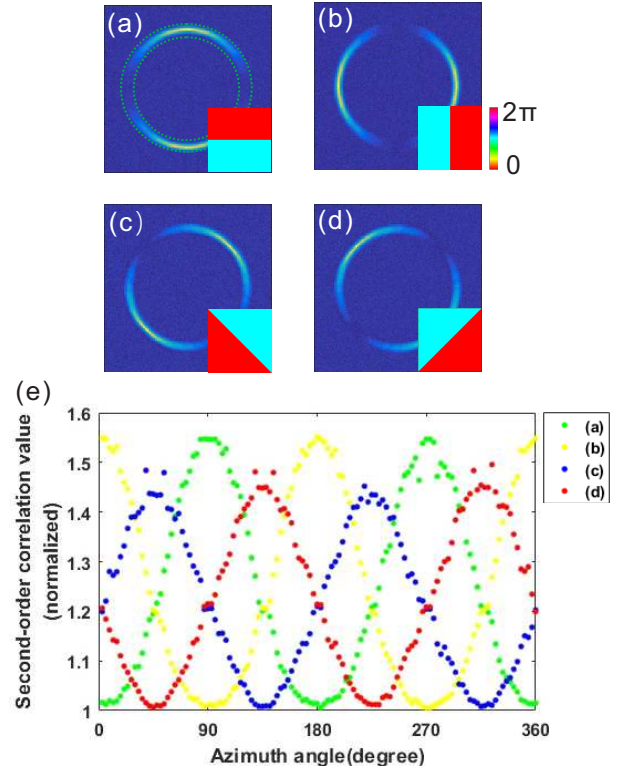}
\caption{(a)-(d) The simulation results of thermal light edge enhancement ghost imaging of the circular phase object with four different holograms orientated at 0,$\frac{\pi}{2}$,$\frac{\pi}{4}$,and $\frac{3\pi}{4}$, respectively. (e) present the second-order correlation values with the azimuthal degree from 0 to $\pi$ in (a)-(d). The dashed lines show the scale in which each point of the curve in (e) are calculated from each local position. }
\label{Fig.3}
\end{figure}

\section{IV. CONCLUSION}
 In summary, we propose a thermal light edge enhanced ghost imaging scheme of phase object. The azimuthal phase filter is placed non-locally to the phase object, which are in the spatially separated light beams. Isotropic and orientational edge enhanced ghost images of the $\pi$-phase step object have been achieved by the phase filter with different OAM topological charges. The Bell-type inequality has also been measured with a circular phase step object and oriented phase filter. The experimental results show that the maximal violation of the Bell inequality is less than 2, which means the classical signature of the thermal light ghost imaging system. Compared to the conventional optical edge enhanced imaging technology, the proposed scenario here can survive with incoherent light source and the edge enhanced image can even be obtained with non-localized phase filter. It may find potential applications in the fields of medical detection, microscopy imaging and remote sensing.



\section{ACKNOWLEDGEMENTS}
\begin{acknowledgments}
This work was supported by the Fundamental Research Funds for the Central Universities and National Innovation and Entrepreneurship Training Program for College Student.
\end{acknowledgments}

\end{document}